\def\spose#1{\hbox to 0pt{#1\hss}}
\def\approxlt{\mathrel{\spose{\lower 3pt\hbox{$\sim$}}
        \raise 2.0pt\hbox{$<$}}}
\def\approxgt{\mathrel{\spose{\lower 3pt\hbox{$\sim$}}
        \raise 2.0pt\hbox{$>$}}}
\def\Msun{\hbox{$\rm\thinspace M_{\odot}$}}

\def\K{{\rm\thinspace K}}
\def\keV{{\rm\thinspace keV}}

\def\Msun{\hbox{$\rm\thinspace M_{\odot}$}}

\def\mdot{\hbox{$\dot{m}$}}
\def\Mdot{\hbox{$\dot{M}$}}
\def\spose#1{\hbox to 0pt{#1\hss}}
\def\approxlt{\mathrel{\spose{\lower 3pt\hbox{$\sim$}}
        \raise 2.0pt\hbox{$<$}}}
\def\approxgt{\mathrel{\spose{\lower 3pt\hbox{$\sim$}}
        \raise 2.0pt\hbox{$>$}}}

\documentstyle[bo99,epsfig]{article}

\title{Coronal heating and \\ emission mechanisms in AGN}

\author{Tiziana Di Matteo\thanks{Chandra Fellow}}
\affil{Harvard-Smithsonian Center for Astrophysics,
60 Garden St., Cambridge, MA 02138; \\tdimatteo@cfa.harvard.edu}
\begin{document}

\maketitle

\begin{abstract}

Popular models for the formation of X-ray spectra in AGN assume that a
large fraction of the disk's angular momentum dissipation takes place
in a hot corona that carries a small amount of the accreting
mass. Here I discuss the formation of a magnetically--structured
accretion disk corona, generated by buoyancy instability in the disk
and the heating of localized flare regions up to the canonical X-ray
emitting temperatures. I also examine the analogy between accreting
disk-coronae and ADAFs and discuss the relevant emission mechanism in
these two accretion models and how observational constraints can allow
us to discriminate between them.

\end{abstract}

\section{Introduction}

Observations of the central regions of AGN show that a significant
fraction of their bolometric luminosity comes out in hard X-rays (from
$\sim 0.1$ ~\keV all the way up to a few 100 \keV) and sometimes up to
1 GeV.

According to the standard paradigm, AGN are powered by accretion onto
their central black hole. An accretion disk around a supermassive
black hole (in an AGN) leads to the production of a strong
optical/ultraviolet continuum, the so--called 'blue bump'. Such a
component is attributed to quasi-blackbody emission (e.g. see Koratkar
\& Blaes 1999 for relevant modifications to the blackbody spectrum for an
accretion disk). The effective absorptive optical depth in a disk is
typically $\tau >> 1$ which implies that photons are close enough to
being in thermal equilibrium with the electrons to produce
a blackbody--like spectrum.  The luminosity of this component scales as
$L \sim \pi r_{g} \sigma T^4$ where $r_g$ is the Schwarzchild
radius. This implies gas temperatures in the disk of the order of
\begin{equation}
 T \sim 5 \times 10^5 L_{44}^{-1/4} \left(\frac{L}{L_{Edd}}\right)^{1/2} \;\; K,
\end{equation}
where $L=10^{44}L_{44}$. $L_{Edd}$ is the Eddington luminosity and the
temperature decreases with increasing luminosity (or increasing black
hole mass).

It is evident from Eqn. 1 that if AGN generated their energy solely by
accretion of matter in thermodynamic equilibrium, the highest
temperatures achieved would be of the order of $10^5 \K$ and
negligible X-ray emission would be expected. Phenomenologically,
therefore, we know that there must be an efficient mechanism for
transferring the energy released in an accretion disk into a plasma
component that is far from thermodynamic equilibrium with the ambient
radiation and that radiates the high energy portion of AGN spectrum.

Although there are many uncertainties concerning how such energy
transfer occurs, we know there must be mechanisms that can sustain the
presence of a very hot plasma near an accretion disk: e.g. the Sun
which has a the surface temperature of only $5500 \K$, is
surrounded by a magnetically-dominated corona with a temperature of
$2-3 \times 10^6
\K$.

Here, I address the issue of why we expect hot electrons to be present
in AGN. I will discuss how AGN coronae formation can be understood as
a direct consequence of the internal dynamics of an accretion disk
where shock-like events (magnetic reconnection and MHD processes) are
responsible for heating the coronal plasma.  I will examine the
relevant radiative processes in AGN that are responsible for the
production of the X-ray emission that we observe.  In particular I
will discuss the relevance of these processes for both AGN coronae and
hot advection-dominated accretion flows (ADAFs) and their relative
importance for different regimes of source luminosities.

\subsection{The X-ray emission} 
Before discussing in more detail the proposed picture of coronae
formation I will briefly review the observed characteristics of the
X-ray emission in AGN and the information that these give when trying
to construct a model for coronae.

Assuming for now the existence of a hot plasma (see Section 2), it is
well established that the X-ray continuum in AGN can be explained by
thermal Comptonization of the soft UV radiation (e.g Haardt \&
Maraschi 1993). There is evidence also that this X-ray continuum
is reprocessed in a cold medium (e.g. the accretion disk) and gives
rise to a reflection bump at around $30 \keV$ and a broad iron, Fe
$K\alpha$ emission line at 6.4 keV.
%--------------------------  figure 1
%\begin{figure}
%\centerline{
%\hbox{~~
%\vspace{-1cm}
%~~
%\psfig{file=refl_spec_chris.ps,width=3.cm,angle=270}
%\vspace{-1cm}
%\psfig{file=julia_var.ps,width=3.cm,angle=270}}
%\vspace{-0.3cm}}
%
%\caption[]{On the left: the effects of X-ray reflection (Reynolds et al. 1996). The dashed line represents the incident continuum with a typical spectrum $F_E \propto E^{-0.9}$. The solid line shows the pure reflection spectrum. The right panel shows ASCA observations of the 1-10 \keV flux from the Seyfert I galaxy
%MGC 6-30-15 (Lee et. al 1999)}
%\end{figure}
%---------------------------------
The presence of these features in the spectrum place constraints on the
geometry of the X-ray emitting region and tell us that the hot plasma
has to be situated above the colder accretion disk. Also, the
different ratios of soft luminosity (attributed to the accretion disk)
to hard X-ray luminosity imply that the hot coronal plasma is not a
slab but consists of localized active regions (e.g. Haardt, Ghisellini
\& Maraschi 1994). This is also consistent with the characteristically
short X-ray variability timescales observed in Seyfert galaxies (as
short as a few hours) which imply that enormous
amounts of energy are released in a very short time in flare-like
events.

Finally, the average X-ray spectra of Seyfert galaxies shows a
high-energy cutoff usually above 100 \keV~which can be reproduced
quite well by models of thermal Comptonization. The absence of
conspicuous electron pair annihilation line indicates that most of the
hot electrons in a corona are thermal. Whatever the processes that
operate in coronae to heat the plasmas are, they do not accelerate a
large number of electrons. Alternatively, mechanisms exists for
efficiently thermalizing the electron population (e.g. Svensson \&
Ghisellini 1998).

\section{Accretion disk coronae}
In recent years significant progress has been made in understanding
accretion disks and how angular momentum transport operates with the
identification of the Balbus--Hawley instabily (e.g. Balbus \& Hawley
1997) for weakly magnatizes disks. Thanks to this fundamental progress,
we can now think more confidently of coronae formation as a direct
consequence of the internal dynamics of an accretion disk; much the
same way the solar corona is thought to be heated by dynamical
processes lower in the Sun's atmosphere.
%--------------------------  figure 3 
\begin{figure}
\vspace{-2cm}
\centerline{\psfig{file=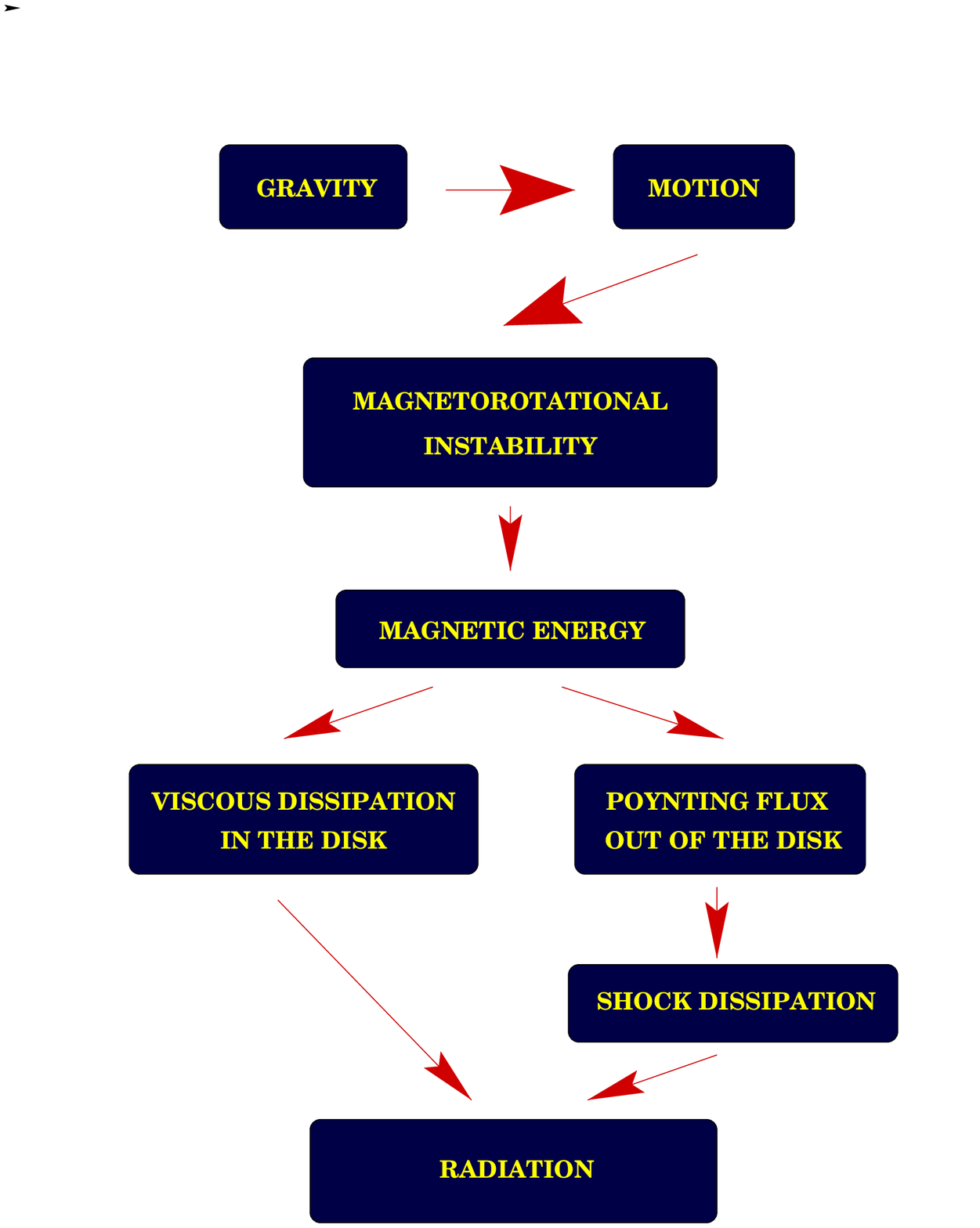,width=8cm}}
\caption[]{Energy flow and angular momentum transport in accretion disks}

\end{figure}
%---------------------------------
More specifically, Balbus \& Hawley have identified an instability in
weakly magnetized accretion flows that is responsible for the
transport of angular momentum. The way such a magneto-rotational
instability works is by producing strong amplification of the seed
magnetic fields and in this way channelling the energy present in the
system into magnetic energy (see Figure 1). The formation of a corona
can be undestood as an efficient way for a disk to saturate the
Balbus-Hawley instability and to dissipate the accretion
energy/angular momentum into particles, which can then radiate it
away. The built-up magnetic energy is dissipated into particles
locally in the disk and partly builds-up strong magnetic flux tubes
leading to a net vertical flux of magnetic energy which inevitably
escapes from the disk to form a magnetically-dominated corona.

The idea that buoyancy of strong flux tubes in the disk and their
expulsion from the disk to form magnetic coronae has been
proposed in the past (Stella \& Rosner 1984, Coroniti 1981, Galeev,
Rosner \& Vaiana 1979), but can only now be integrated in a deeper
understanding of accretion phenomena.

\subsection{Coronal heating: magnetic reconnection}
Within the context of such a model, the question of how the coronal
plasma heats up to X-ray emitting temperatures can be assessed.
Such coronae (e.g. ensembles of flux tubes) contain a very
small amount of mass and are magnetically dominated. By definition, the
magnetic flux tubes become buoyant when $\beta\sim 8\pi P/B^2
\approxgt 1$ where $B$ is the magnetic field strength and $P$ the gas
pressure in the disk. The typical speed of the rising flux tubes is
then given by their Alven speed e.g. $V_{A} = B/\sqrt{4\pi \rho}
\approxgt c_{\rm s}$ and is by definition always larger then the relevant
sound speed ($c_{\rm s}$) implying (in a simple view) that the buoyant
magnetic energy has to be dissipated in shocks. So, whereas the core
of the disk is usually dominated by subsonic turbulence the coronal
gas above the disk is, inevitably, supersonic.

More realistically we would expect this energy to be dissipated in
shocks in reconnection sites where strong impulsive heating occurs
when magnetic field lines are brought together. Reconnection can occur
either `spontenously' in a given magnetic loop or can be `driven' when
more than one magnetic tubes are brought together.  A reconnection
site is thought to be a collection of particle acceleration and
heating (e.g. direct Joule heating near X-point, slow shock
acceleration, Fermi magnetic mirroring in turbulent outflows,
conduction, downstream fast shocks etc..) but the
detailed physics of how it occurs is still an unsolved MHD problem
(for the case Petschek reconnection).

Although the general physical picture of accretion disk coronae
described above provide us with an understanding of why we expect to
find hot plasmas above accretion disks, there remain many
uncertainties. These include the question of which pressure is
relevant for magnetic field amplification and buoyancy.  It is not clear
whether magnetic fields build up to equipartition such as $B^2/8\pi
\sim P_{tot}$ or $B^2/8\pi \sim P_{gas}$
Also, it is uncertain what fraction of the magnetic energy is dissipated
into $e^{-}$ and $p$. It is clear that when energy dissipation
occurs one needs to treat the plasma as a 2-temperature medium:
different wave-particle interactions will heat electrons and protons
differently. One can construct 2-T AGN coronae if the protons contain
most of the energy (Di Matteo, Blackman \& Fabian 1997), but no
clear-cut arguments can be made to support their plausibility over one
tempearature models. The same problems exist in the case of ADAF
plasma where the 2-T condition is a crucial assumption but, at this
stage, yet to be proven.  Coronal plasmas are often collisionless. It
is not clear, therefore, whether electrons are thermalized or
dissipative processes cannot accelerate particles efficiently. In
other words the importance of direct heating versus acceleration in
either coronal or ADAFs plasma, cannot be determined.  In AGN coronae
or ADAFs, $V_{A}$ can approach $c$, and one should really consider the
effects of relativistic MHD. Such effects are usually not taken into
account.

Important recent results from numerical simulations (Miller \& Stone
1999) do indeed show the formation and heating of magnetized coronae
above accretion disks. In particular, Miller \& Stone have shown that
when weak $B$ fields are amplified in the disk via MHD turbulence
driven by the Balbus--Hawley instability some of the magnetic energy is
dissipated locally but a good fraction escapes due to buoyancy and
forms a strongly magnetized corona above the disk. Most of the energy
in their simulations is dissipated at a few scale heights above the
disk, and strong shocks are continuously produced making the corona hot
up to X-ray emitting temperatures.  Their results on the impulsive
heating of coronal plasmas, are in accordance with simple analytical
estimates (Di Matteo 1998) on the occurrence of an ion--acoustic
instability, associated with slow shocks in Petschek magnetic
reconnection in flare-like events in a magnetically-dominated
corona. The occurrence of an ion--acoustic instability, associated
with slow shocks in Petschek magnetic reconnection, can be shown to
result in a violent release of energy and heat the coronal plasma to
canonical X-ray emitting temperatures (of a few $\times 10^9$K).

\section{Emission mechanisms}
In the previous sections I have discussed the vertical structure of
an accretion disk and how its internal dynamics can lead to the
formation of a highly-dynamic, magnetically dominated and heated
corona. 

Both in AGN coronae and in ADAFs (also magnetized and with hot $\sim
10^9 \K$ electrons; see Narayan, Quataert \& Mahadevan 1999 for a
recent review) the relevant interactions and relative emission
mechanisms are: particle-photon $\rightarrow$ Compton processes;
particle-magnetic field $\rightarrow$ cyclo/synchrotron emission and
particle-particle $\rightarrow$ bremsstrahlung emission.

Inverse-Compton scattering of disk photons off the hot electrons is
usually the dominant process in most AGN.  The importance of
Inverse-Compton processes scales $ \propto U_{rad} \exp(y)$ where $y
\sim 4 (kT/m_ec^2) \tau $, $\tau = n_e \sigma_T r$ and the energy density
 $U_{rad} \sim L/(R^2c)\tau$ is usually attributed to the external
soft photon flux coming from the disk. Bremsstrahlung instead scales
as $n_e^2 T^{-1/2}$, where $n_e$ is the electron number density, and
dominates IC only in very low luminosity objects e.g. 
\begin{equation}
IC > BREM \rightarrow \frac{L}{L_{Edd}} > \frac{10^{-5}}{\sqrt{\theta}} r_s
\end{equation}
(see also Section 3.2), where $\theta$ is the dimensionless electron
temperature.

\subsection{Synchrotron emission and Comptonization in coronae and ADAFs}
Both in the case of an AGN corona or an ADAF, magnetic fields are
close to their equipartition values and synchrotron emission should be
taken into account.

In both cases electrons are considered to be thermal. Thermal
synchrotron is heavily self-absorbed up to a frequency $\nu_{s}
\propto T^2 B$. Equipartition arguments (in the case of a supermassive
black hole with $M \sim 10^7 \Msun$) imply values of $B \sim
P_{gas} + P_{rad} \sim 10^{3-5}$ Gauss and for canonical corona
temperatures of $10^9 \K$, synchrotron emission peaks in the
Infrared/Optical bands (Di Matteo, Celotti \& Fabian 1997; see Figure
2a).

%--------------------------  figure 4 
\begin{figure}
\centerline{\psfig{file=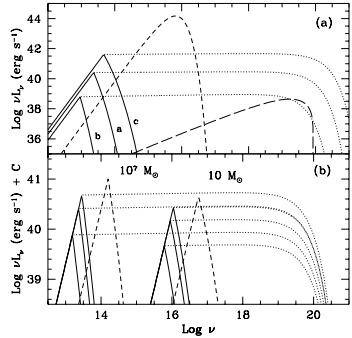,width=4.5cm}}
\vspace{-0.3cm}
\caption[]{(a) The solid line is the synchrotron component, the dotted
line its Comptonization, small dashed line the blackbody component
form the accretion flow. (b) The shift in frequency of the
synchrotron peak between supermassive and galactic black holes.}

\end{figure}
%---------------------------------
The synchrotron soft photon flux is Inverse Compton scattered up to
X-ray energies by the hot electrons (dotted line in Figure 2a). In
most cases though, synchrotron Inverse Compton does not dominate the
X-ray emission because the energy density due to the soft disk photon
field dominates the scattering. Due to the high self-absorption, the
synchrotron energy density $U_{syn} < B^2/8\pi$ which, given the
equipartion arguments, implies $U_{syn} < U_{disk}
\sim P_{rad}$ (Fig. 4a) and Comptonization of the soft disk photons
dominates the X-ray emission. Given the strong dependences of thermal
synchrotron emission on both temperature and $B$, and the very
dynamical structure of the corona, estimates of an 'average' $T$ and
$B$, which are usually employed in these calculations are likely to be
unrealistic. As shown by the above relations, the importance of
synchrotron and its Inverse Compton might be highly enhanced if flares
are at different temperatures and some are hotter and/or with higher
magnetic fields than the values usually assumed from global
arguments. It is plausible that a non-thermal population of particles
could be present which would also significantly enhance the
synchrotron and its IC component but this has not been taken into
account in current models).

In contrast, in an ADAF the synchrotron photons are, in most cases,
the only source of soft photons for Comptonization (even if the ADAF
is matched to a thin disk at large distances, as in models by Esin et
al. 1997; Quataert et al 1999 its contribution is negligible; e.g. see
Figure 3).

Comptonization of the synchrotron component in an ADAF can explain the
observed X-ray emission in some low-luminosity AGN. Figure 3
shows the case for M81 and NGC 4579 both of which have an estimated
mass for the central black hole, detected hard X-ray emission, and
optical/UV emission too low to allow for the presence of a
geometrically thin, optically thick accretion disk close to the black
holes (Quataert et al. 1999).
% ------------------- figure 5 
\begin{figure}
\hbox{
\centerline{
\psfig{figure=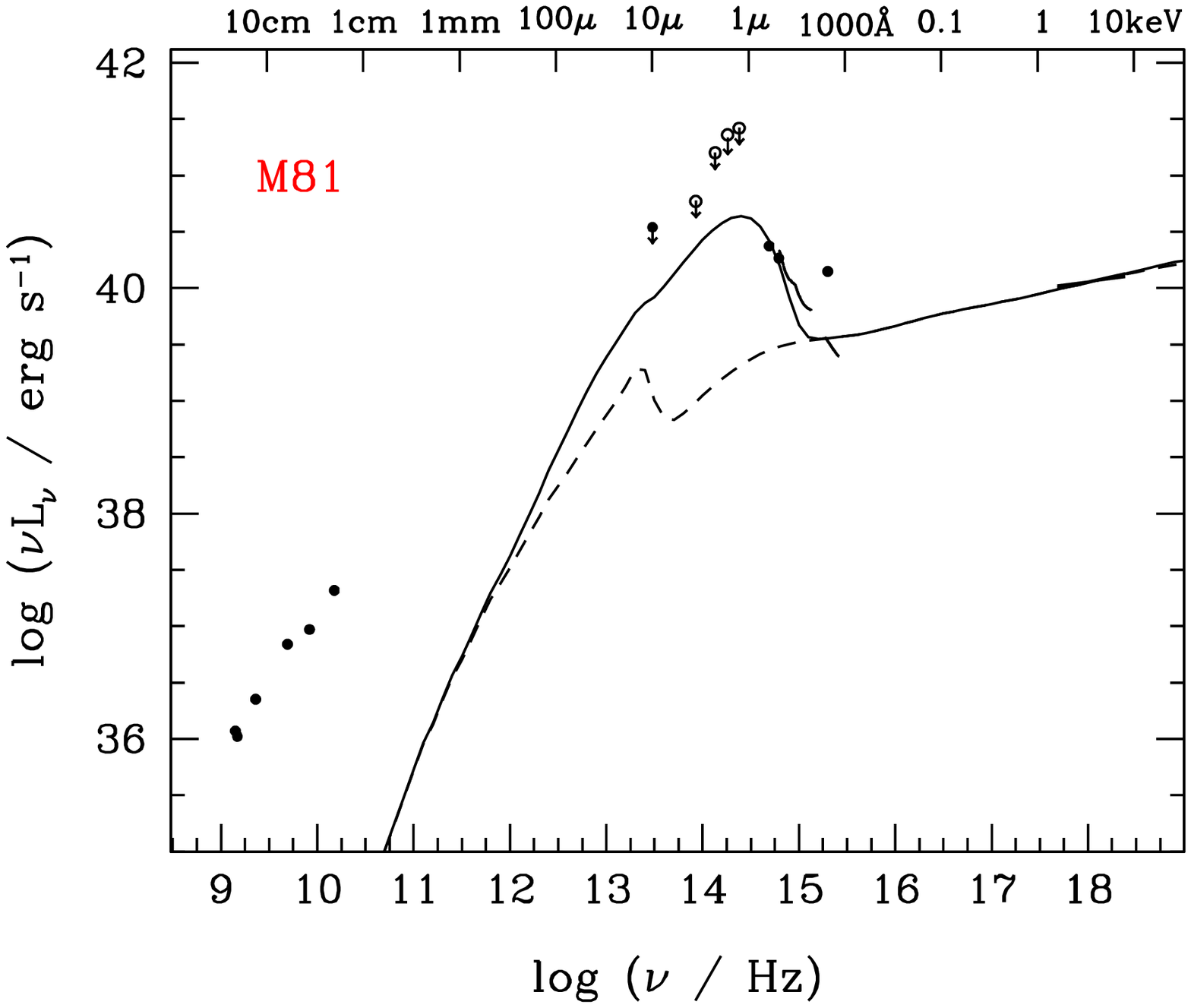,width=5cm}
\hspace{-0.7cm}
\psfig{figure=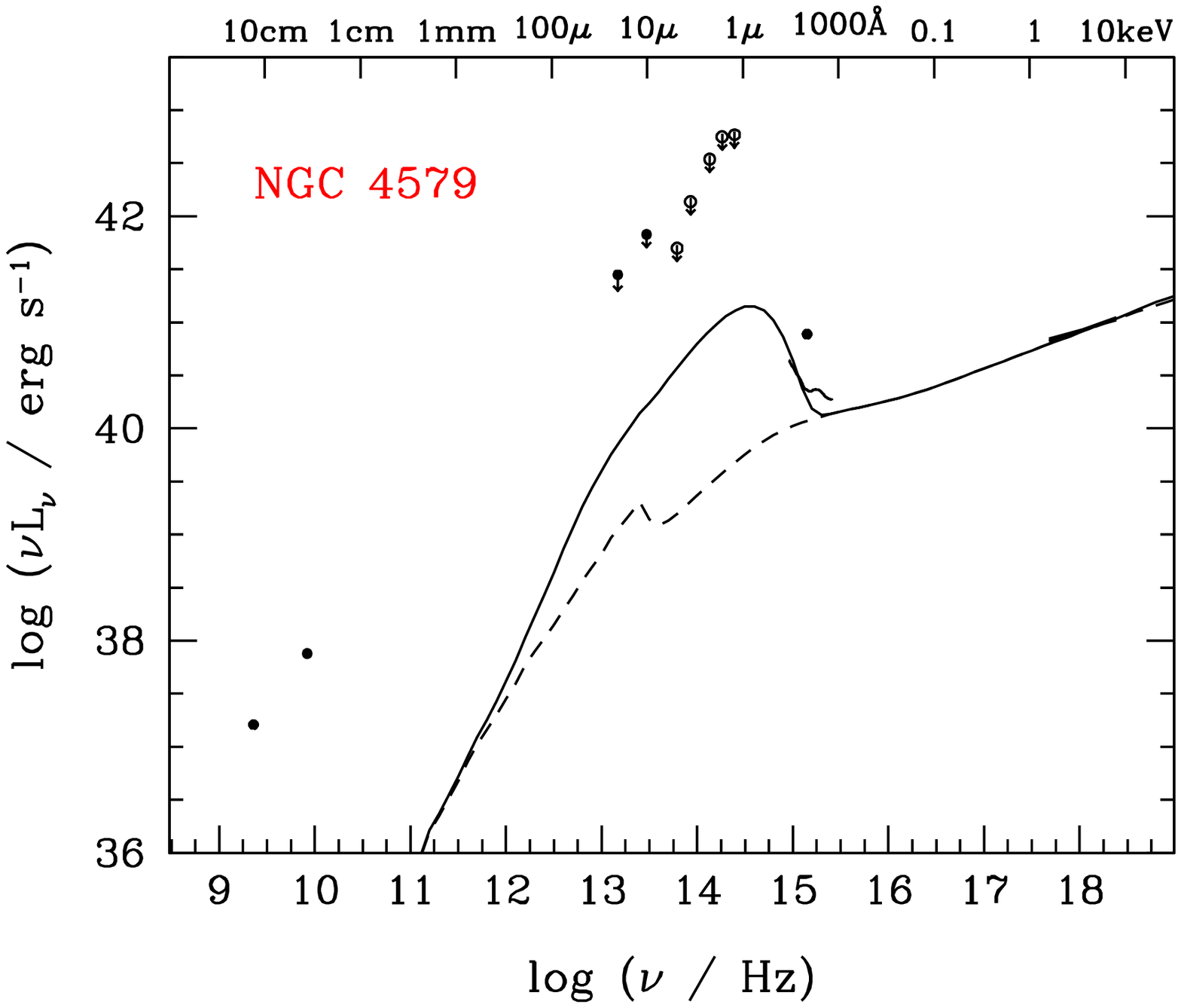,width=5cm}
\hspace{-0.7cm}
\psfig{figure=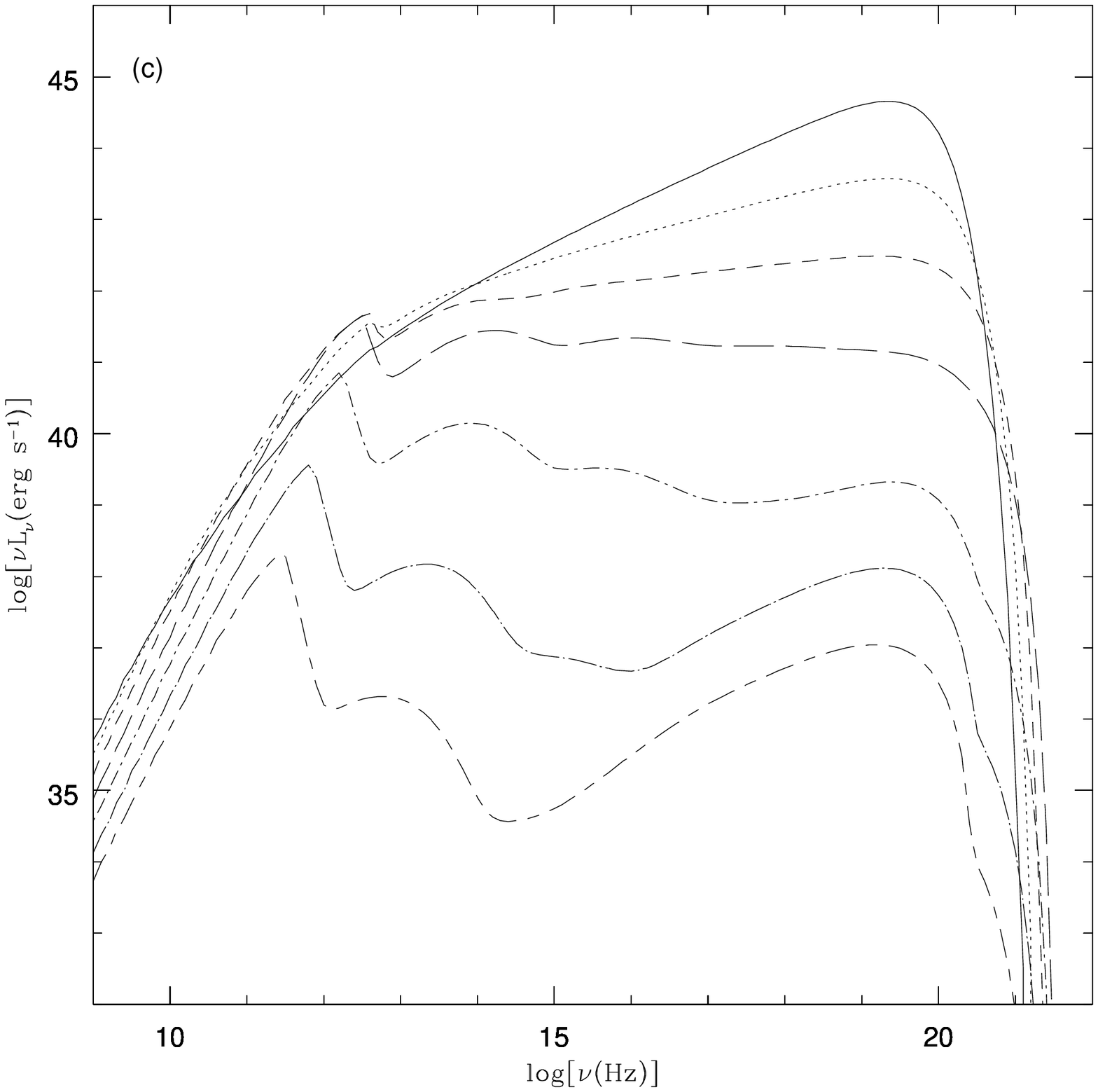,width=4.7cm}
}}
\vspace{-0.3cm}
\caption[]{ Model for M81 NGC 4579 in which a thin disk is truncated
at $r \sim 100 r_{S}$, inside of which there is an ADAF. The solid
line shows the total "disk + ADAF" emission, while the dashed line
shows the ADAF contribution (Quataert et al. 1999). On the rightmost
panel the spectra of ADAFs for a $10^9 \Msun$ black hole with $\mdot$
decreasing by about 3 orders of magnitude from the top curve to the
bottom one. The high energy spectrum changes from Comptonized
synchrotron to bremsstrahlung.}
\end{figure}
%-------------------------
In general, in a standard ADAF, Comptonization becomes important for
$\mdot \approxlt \mdot_{crit}$ above which the hot flow cannot
exist. In the high $\mdot$ regime considered here, the characteristic
electron scattering optical depth $\tau$ of the ADAF becomes of order
unity since $\tau \propto \mdot$.  As $\tau$ decreases with decreasing
$\mdot$, bremsstrahlung becomes the dominant process (see Figure 3).
%%--------------------------  figure 6 
%\begin{figure}
%\centerline{\psfig{file=adaf_spec.ps,width=4cm}}
%\vspace{-0.2cm}
%\caption[]{Spectra of ADAFs for a $10^9 \Msun$ black hole with $\mdot$
%decreasing by about 3 orders of magnitude from the top curve to the
%bottom one. The high energy spectrum changes from Comptonized synchrotron
%to bremsstrahlung.}
%
%\end{figure}
%---------------------------------
\subsection{Bremsstrahlung emission in elliptical galaxy nuclei}

Equation 2 shows that bremsstrahlung emission can only be important in
 sources with very low luminosities (or low radiative efficiencies).
 The nuclear regions of elliptical galaxies provide excellent
 environments in which to study the physics of low-luminosity
 accretion.  There is now strong evidence, from high-resolution
 optical spectroscopy and photometry, that black holes with masses of
 $10^8-10^{10} \Msun$ reside at the centers of bulge dominated
 galaxies, with the black hole mass being roughly proportional to the
 mass of the stellar component (e.g. Magorrian et al. 1998). X-ray
 studies of elliptical galaxies also show that they possess extensive
 hot gaseous halos, which pervade their gravitational
 potentials. Given the large black hole masses inferred, some of this
 gas must inevitably accrete at rates which can be estimated from
 Bondi's spherical accretion theory. Such accretion should, however,
 give rise to far more nuclear activity (e.g. quasar-like
 luminosities) than is observed, if the radiative efficiency is as
 high as 10 per cent (e.g. Fabian \& Canizares 1988), as is generally
 postulated in standard accretion theory.

Accretion with such high radiative efficiency need not be universal,
however. As suggested by several authors (Rees et al. 1982; Fabian \&
Rees 1995), the final stages of accretion in elliptical galaxies may
occur via an advection-dominated accretion flow (ADAF; Narayan \& Yi
1995, Abramowicz et al. 1995) at roughly the Bondi rates. Within the
context of such an accretion mode, the quiescence of the nuclei in
these systems is not surprising; when the accretion rate is low, the
radiative efficiency of the accreting (low density) material will also
be low. Other factors may also contribute to the low luminosities
observed. As discussed by Blandford \& Begelman (1999; nad emphasized
observationally by Di Matteo et al.  1999a), and shown numerically by
Stone, Pringle \& Begelman (1999), winds may transport energy, angular
momentum and mass out of the accretion flows, resulting in only a
small fraction of the material supplied at large radii actually
accreting onto the central black holes.

If the accretion from the hot interstellar medium in elliptical
galaxies (which should have relatively low angular momentum) proceeds
directly into the hot, advection-dominated regime, and low-efficiency
accretion is coupled with outflows (Di Matteo et al. 1999a), the
question arises of whether {\it any} of the material entering into the
accretion flows at large radii actually reaches the central black holes.  
The present observational data generally provide little or no evidence for
detectable optical, UV or X-ray emission associated with the nuclear
regions of these galaxies.

The discovery of hard X-ray emission from a sample of six nearby
elliptical galaxies (Allen, Di Matteo \& Fabian 1999), including the
dominant galaxies of the Virgo, Fornax and Centaurus clusters (M87,
NGC 1399 and NGC 4696, respectively), and NGC 4472, 4636 and 4649 in
the Virgo cluster, has important implications for the study of
quiescent supermassive black holes. The ASCA data for all six sources
provide clear evidence for hard, power-law emission components, with
photon indices in the range $\Gamma = 0.6-1.5$ and intrinsic $1-10$
keV luminosities of $2 \times 10^{40}-2 \times 10^{42}$ erg s$^{-1}$
(Allen et al. 1999). This potentially new class of accreting X-ray
source has X-ray spectra significantly harder than Seyfert nuclei and
bolometric luminosities relatively dominated by their X-ray emission.

We argue that the X-ray power law emission is most likely to be due to
accretion onto the central supermassive black holes, via low-radiative
efficiency accretion (Allen et al. 1999, Di Matteo et al. 1999b).

%------------------Figure --------------------------
\begin{figure}
\centerline{
\hbox{
\hspace{-0.3cm}
\psfig{file=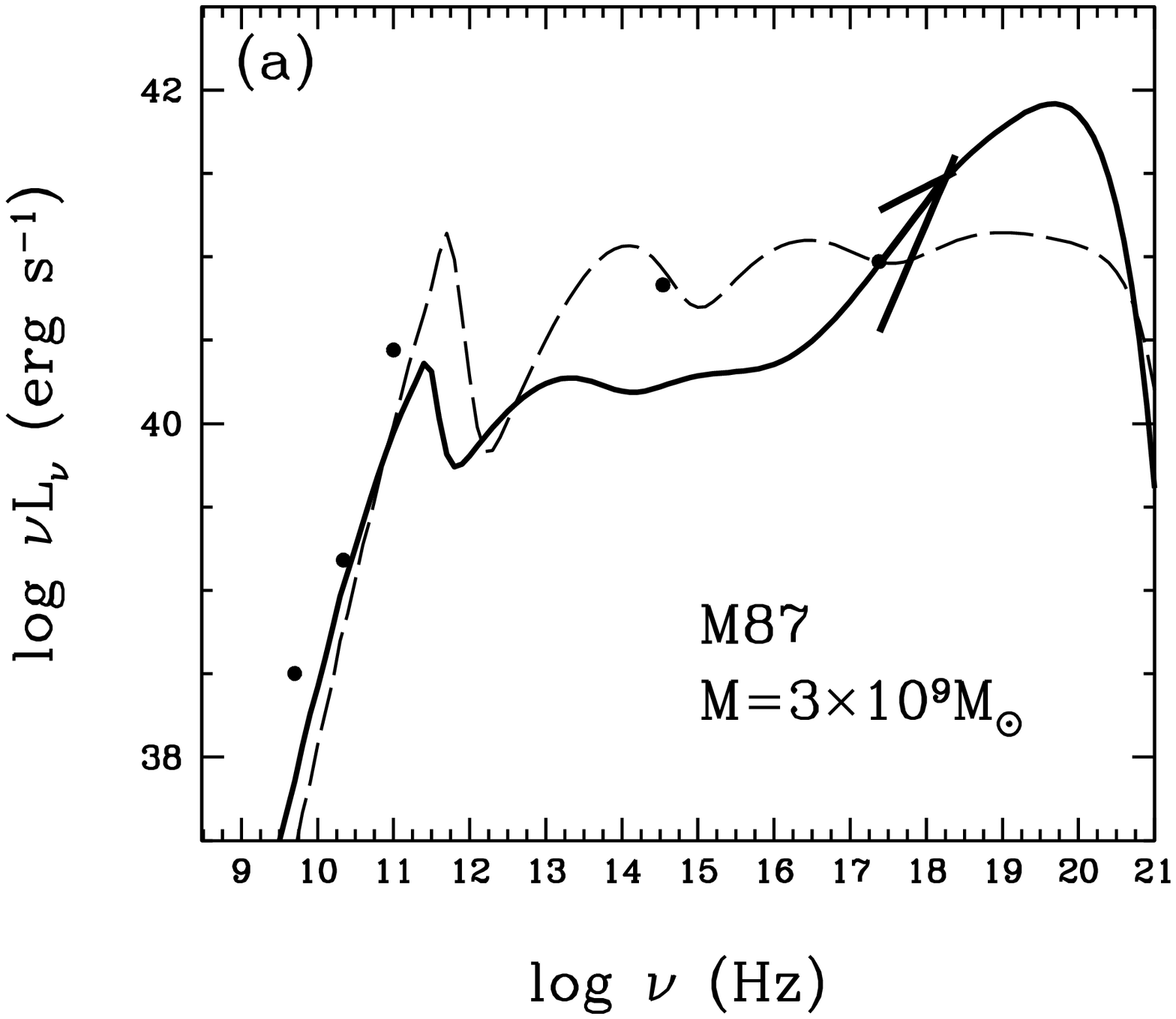,width=5cm}
\hspace{-0.7cm}
\psfig{file=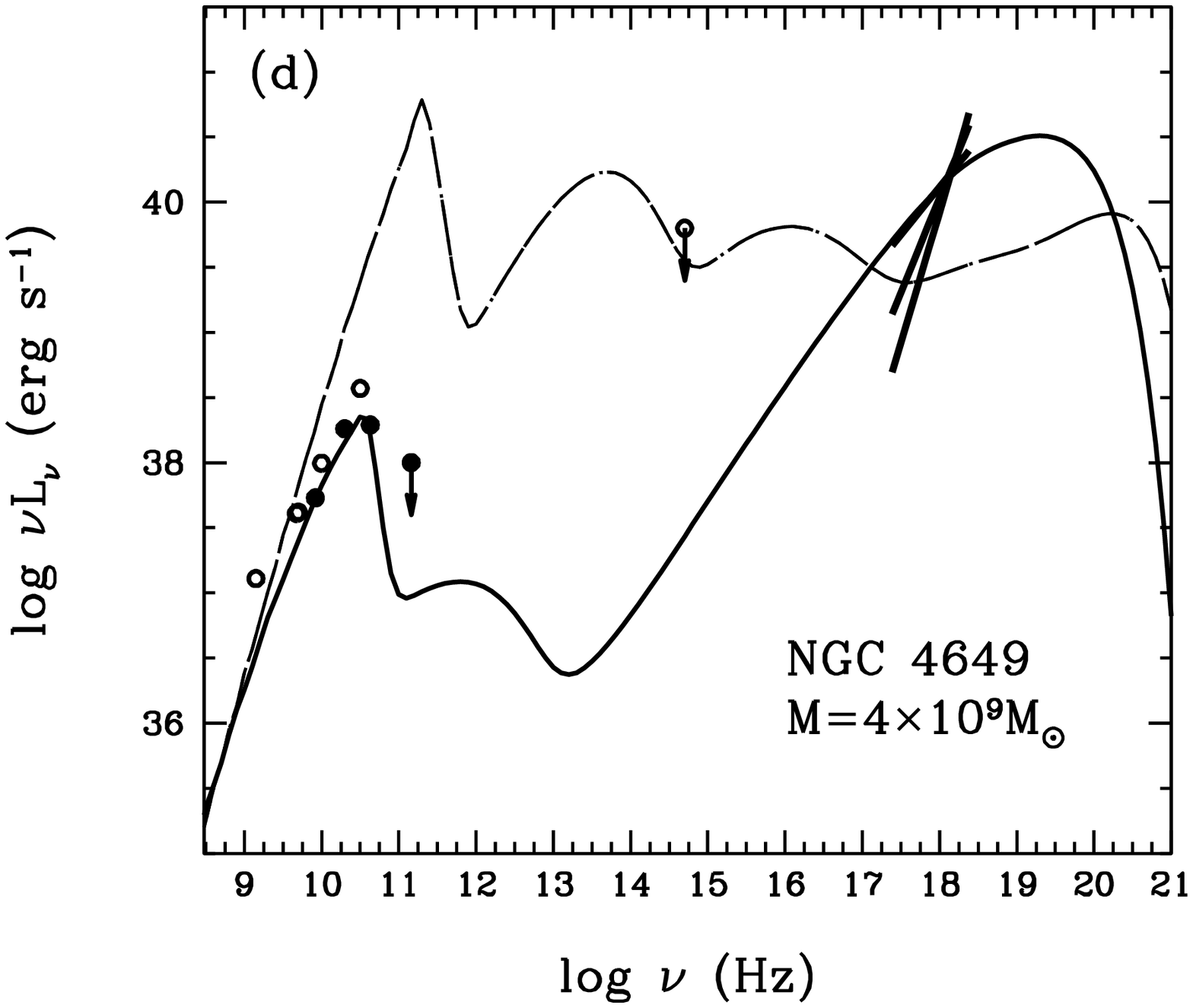,width=5cm}
\hspace{-0.5cm}
\psfig{file=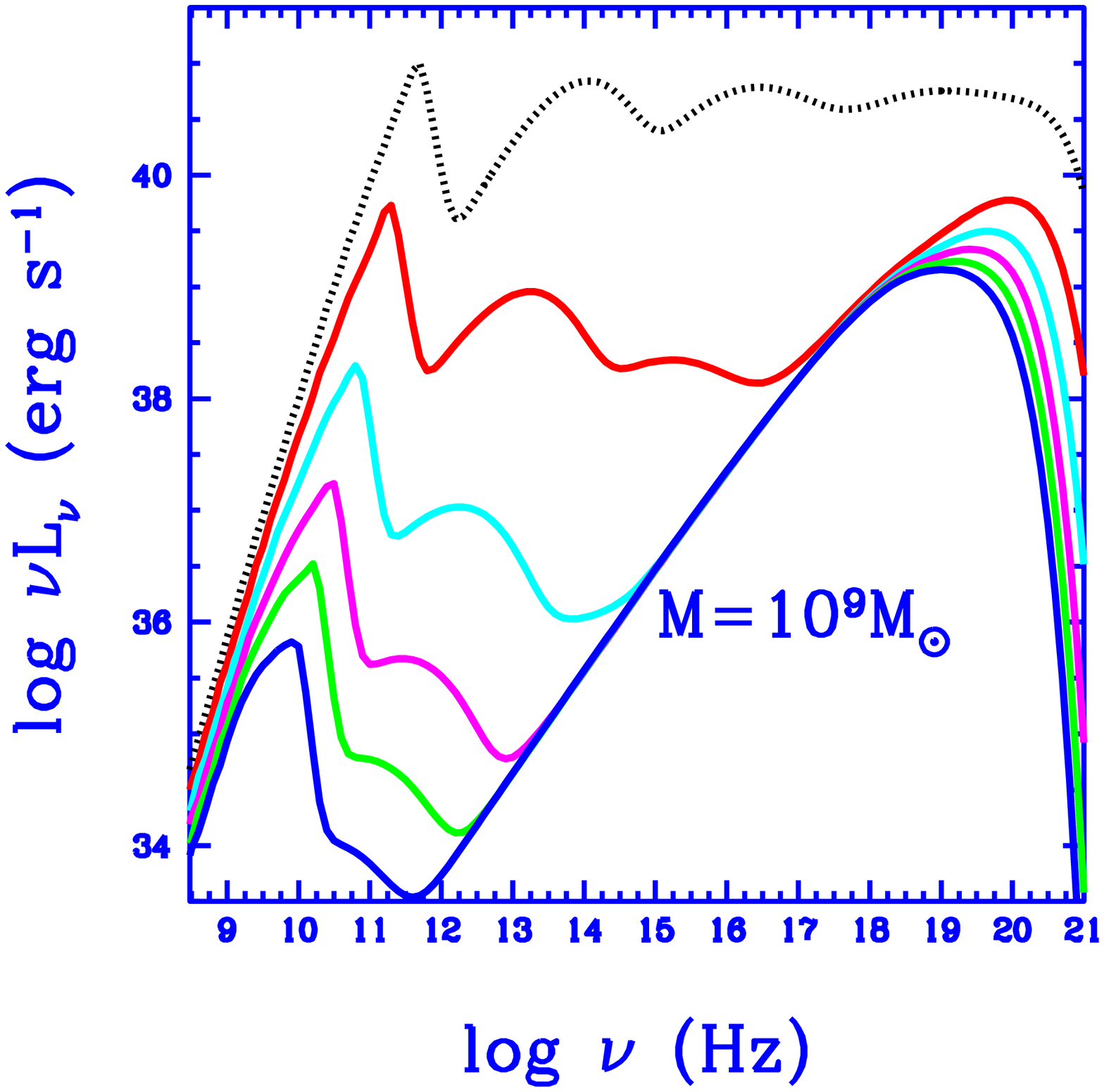,width=5cm}}
}
\vspace{-0.2cm}
\caption{ Spectral models calculated for ADAF with outflows and
without outflows (dashed lines) are shown for two representative cases
(for the other objects see Di Matteo et al. 1999b).  The solid dots
are the best constraints on the core emission. The thick solid lines
the slopes and fluxes obtained from the ASCA analysis.  The leftmost
panel shows explicitly the effects of winds on the spectra of ADAFs,
The mass outflow scales as $\Mdot = \Mdot_{out}\left(\frac{R}{R_{\rm
out}}\right)^p $. Models are calculated for the same $\Mdot_{out}$ and
$R_{\rm out}$ with $p$ increasing from 0 to 1 (in steps of 0.2) from
the top dashed curve to the lower one}
\end{figure}
%---------------------------------------------------

The broad band spectral energy distributions for these galaxies, which
accrete from their hot gaseous halos at rates comparable to their
Bondi rates, can be explained by low-radiative efficiency accretion
flows in which a significant fraction of the mass, angular momentum
and energy are removed from the flows by winds.  The observed
suppression of the synchrotron components in the radio band (Di Matteo
et al. 1999a; excluding the case of M87) and the systematically hard
X-ray spectra, which are interpreted as thermal bremsstrahlung
emission, support the conjecture that significant mass outflow is a
natural consequence of systems accreting at low-radiative efficiencies
(see the representative cases of NGC 4649 and M87 in Figure 4 and for
all of the objects Di Matteo et al. 1999b).

The presence of outflows in the hot flows suppresses completely the
importance of Comptonization in ADAF flows and bremsstrahlung becomes
(irrespectively of the accretion rate, c.f. Figure 3) the dominant
X-ray emission mechanism. A representation of the effects on the ADAF
spectra of outflows is shown in Figure 4.

\begin{acknowledgements}
TDM acknowledges support for this work provided by NASA 
through Chandra Fellowship grant number PF8-10005 awarded by the Chandra
Science Center, which is operated by the Smithsonian Astrophysical 
Observatory for NASA under contract NAS8-39073.
\end{acknowledgements}


\begin{references}

\ref Abramowicz M.A., Chen X., Kato S., Lasota J.P., Regev O., 1995, ApJ,
 438, L37

\ref Allen, S.W., Di Matteo, T., Fabian, A.C., 1999, MNRAS, in press

\ref Balbus, S. A., Hawley, J. F., in Accretion Processes in Astrophysica
l Systems: Some Like it Hot! Eighth Astrophysics Conference, College Park, MD, 
October 1997. Edited by Stephen S. Holt and Timothy R. Kallman, AIP Conference
Proceedings 431., p.79

\ref Blandford R.D., Begelman M.C., 1999, MNRAS, 303, L1

\ref Coroniti F.V., 1981, ApJ, 244, 587 

\ref Di Matteo T., 1998, MNRAS, 299, L15

\ref Di Matteo T., Blackman E.G., Fabian A.C., 1997, MNRAS, 291, L23

\ref Di Matteo T., Celotti A., Fabian A.C., 1997, MNRAS, 291, 805

\ref Di Matteo, T., Fabian, A.\,C., Rees, M.\,J., Carilli, C.\,L., Ivison, R.\,J.\, 1999a, MNRAS, 305, 492 

\ref Di Matteo, T., Quataert, E., Allen, S.\,W., Narayan, R., Fabian, A.\,
C., 1999b, MNRAS, in press

\ref Fabian A.C., Canizares C.R., 1988, Nature, 333, 829

\ref Fabian A.C., Rees M.J., 1995, MNRAS, 277, L55

\ref Ghisellini, G., Haardt, F.,Svensson, R. 1998, MNRAS, 297, 348

\ref Galeev A.A., Rosner R., Vaiana G.S., 1979, ApJ, 229, 318

\ref  Haardt F., Maraschi L., ApJ, 1993, 413, 507

\ref Haardt F., Maraschi L., Ghisellini G., 1994, ApJL, 432, L92

\ref Koratkar A., Blaes O., 1999, Publ. Astron. Soc. of Pacific, 111, 1

\ref Magorrian J. et al., 1998, AJ, 115, 2285

\ref Lee J.C,, Fabian A.C., Reynolds C.S., Brandt W.N., Iwasawa K., 1999, MNRAS, submitted

\ref Miller K. A.,  Stone J.M., 1999, ApJ, submitted 

\ref Narayan R., Yi I., 1995, ApJ, 444, 231

\ref  Narayan R., Mahadevan R., Quataert E., 1998, Theory of
Black Hole Accretion Disks, edited by Marek A. Abramowicz, Gunnlaugur
Bjornsson, and James E. Pringle. Cambridge University Press, 1998.,
p.148


\ref Quataert E., Di Matteo T., Narayan R., Ho L., 1999, ApJL,  525, L89

\ref Rees M.J., Phinney E.S., Begelman M.C., Blandford R.D., 1982, Nature
, 295, 17

\ref Reynolds C.S., Fabian A.C., Inoue H., MNRAS, 276, 1311

\ref Stone J.M., Pringle J.E., Begelman M.C., 1999, MNRAS, in press

\ref Stella L., Rosner R., 1984, ApJ, 277, 312

\end{references}
\end{document}